# Evidence of Donor Bias in Chicago Police Stops

Working Paper

April 17, 2025

Angela Zorro Medina[1]    David Hackett[2]    Devin Green[3]    Robert Vargas[4]

## Abstract

This study provides the first empirical evidence that private donations to police departments can influence officer behavior. Drawing on the psychology of reciprocity bias, we theorize that public donations create social debts that shape discretionary enforcement. Using quasi-experimental data from Chicago, we find that after 7-Eleven sponsored a police foundation gala, investigatory stops, particularly of Black pedestrians, increased around its stores. These findings reveal a racialized pattern of donor bias in policing and call into question the consequences of private donations to public law enforcement.

[1] Corresponding author ap.zorro@utoronto.ca, Centre for Criminology & Socio-Legal Studies-Faculty of Law, University of Toronto
[2] Justice Project, University of Chicago
[3] Department of Political Sciences, University of Chicago
[4] Department of Sociology, University of Chicago.

**Introduction**

Private donations to public police departments have long raised questions about their influence on law enforcement behavior. Policing is often framed as a public service —an institution that enforces laws impartially to maintain order and safety. Yet, behind the scenes, a different economic reality is increasingly at play: thousands of police departments in the United States and Canada receive hundreds of millions of dollars annually in private donations from businesses, industry groups, and even individuals to fund operations, equipment purchases, or community engagement work (Walby, Lippert, and Luscombe 2018; Yamawaki Shachter et al. 2025). Scholars of policing have cautioned that corporate gifts to police may create biases or de facto obligations, generating transparency problems and a private co-optation of a public service (Grabosky 2007; Walby et al. 2018). However, to date, the evidence linking private donations to changes in police behavior is very limited.

Why might private donations influence police behavior? Donations can be seen as a form of gift exchange that, while legal, can often create pressures (implicit or explicit) to return the favor (Fong 2023). Psychological scholars argue that cooperation is a critical component of human behavior, and reciprocity plays a key role in maintaining cooperation in contexts where people expect to encounter others repeatedly (Sakaiya et al. 2013). According to the economics of reciprocity, humans give gifts expecting a material benefit from their actions, and the gift recipient has incentives to return the favor even if it is costly for them and yields neither present nor future material rewards (Fehr and Gächter 2000a, 2000b). Evidence of this positive reciprocity has been documented in many trust or gift exchange experiments, and results have been consistent independent of the amount of money at stake (Fehr and Gächter 2000a, 2000b).

We draw upon the psychological literature on "reciprocity bias" to theorize the effects of private donations to police on officer behavior. Recent studies have shown that expectations of reciprocity interfere with individual moral perception and judgment (Zhang et al. 2025). Thus, we argue that private donations to police may influence individual officer behavior by creating what we call a "donor bias." Donor bias has two components: social debt, or an officer's information about benefits from a donor, and reciprocity, a conscious or unconscious feeling that those benefits should be rewarded. According to the reciprocity bias theory, officers' awareness about a donation might trigger reciprocity, affecting officers' perception of fairness, especially when the unfair situation does not personally harm the beneficiary.

To test our theory, we use data on corporate donations to the Chicago Police Department (CPD) and police investigatory stop data between 2016 and 2020. We identify Chicago Police donors that have physical locations across the city to evaluate whether police officers changed their behavior after a donation occurred. We focused on donations through public sponsorship to the Chicago Police Foundation's (CPF) True-Blue Gala (TBG) since it allows us to test the effect of an advertised donation and makes it easy for police officers to identify the sponsor as a police benefactor. Donations to the TBG are publicized on the CPF and CPD's social media pages, as well as in the Fraternal Order of Police's magazine. Additionally, CPD officers attend the gala each year—either as guests or to provide demonstrations for non-police attendees. Attending officers are exposed to information about the gala's donors and have stated that they share it with their peers on the force.

Testing our theory on the entire pool of donors to the TBG was difficult due to the dearth of publicly available data on officer behavior. For example, Citadel LLC founder and CEO Kenneth Griffin has donated millions of dollars to the Chicago Police Department, which may have shaped officer perceptions or behavior toward the company. Citadel LLC's headquarters, however, was in a downtown Chicago office building with low crime. If Citadel LLC were to receive any indirect reciprocity from the police for their donations, it would likely appear through officers with knowledge of individual Citadel employees or through officers' overall attitudes toward the hedge fund. Empirically examining such an effect would require detailed data on officer behavior and perceptions that simply do not exist for independent researchers.

However, a donation from the 7-Eleven retail chain provided an ideal case to test our hypothesis on the effects of donations to police due to the wide distribution of their stores across the city, and the fact that 7-Eleven only sponsored the police gala once during our period of study. Using a pre-registered quasi-experimental design and unique block-level data, we leverage the temporal timing of the donation (November 2017) and the geographic distribution of 7-Eleven stores. We compared blocks with 7-Eleven stores with blocks with major intersections (after making sure that these two groups had comparable crime rates and socio-demographic conditions). We estimate whether changes in police investigatory stop patterns by race occurred after 7-Eleven sponsored the TBG in November 2017.[5]

We estimate 7-Eleven donated $10,000 U.S. dollars as part of their TBG sponsorship. As previous literature has shown, individuals tend to reciprocate gifts independently of the amount, especially when there is the possibility of getting future benefits from their benefactors, like in the case of stable relations of corporate donations to police departments.

Our estimations assessing the effect of small donations on police stop patterns are a critical first step in the systematic evaluation of gift-giving to police departments. Systematic reforms in this area have often focused on illegal cooptation of policing services, leaving the legal gift-giving systems untouched as these have been perceived as unproblematic. Yet, our results show that a reciprocity pattern occurs after a donation, even if it is small, takes place.

---

[5] The preregistered analysis plan is available on the Center for Open Science Achieve at https://osf.io/ye9uk/.

**Results**

Table 1 shows the main findings of our aggregated estimations assessing the effect of 7-Eleven donation on police investigatory stops. After 7-Eleven became the sponsor of the TBG, Blocks with a 7-Eleven within 330 ft experienced a generalized increase in all investigatory police stops. These aggregated results suggest that the Chicago Police Department responded to the 7-Eleven donation, making 389 additional investigatory stops around the convenience stores of their benefactor. As we anticipated, pedestrian stops drive the effect, and no statistically significant effect is observed in vehicular stops. 7-Eleven stores are walk-up accessible, often located in high-foot traffic areas, open late, and frequented by pedestrians. This suggests that changes in police investigatory stops are consistent with a reciprocity intention to benefit their benefactor. Chicago experienced an increase of 272 additional pedestrian investigatory stops around 7-Eleven stores in the following two years after the donation.

The increase in investigatory stops is not evenly distributed across different racial groups. Table 1 shows that police officers in Chicago made 272 more investigatory stops of black people and 261 more investigatory stops of black pedestrians following two years after the donation in blocks with 7-Elevens after November 2017. While an increase is also observed for hispanics (51 additional stops) and whites (59 additional stops), the donation effect was substantially larger for black individuals. While these subgroup models were estimated separately, the magnitude of the effect for blacks was more than four times that for whites, suggesting a potentially racialized pattern in how police activity responded to the private donation.

Table 1: Difference-in-Difference Results 7-Elevens vs Major Intersections

|  | All Races | Black | White | Hispanic |
|---|---|---|---|---|
| **All Investigatory Stops** | | | | |
| Donation | 0.620*** | 0.433*** | 0.094*** | 0.081** |
|  | (0.131) | (0.100) | (0.026) | (0.028) |
| **Pedestrian Stops** | | | | |
| Donation | 0.577*** | 0.415*** | 0.089*** | 0.064** |
|  | (0.118) | (0.093) | (0.024) | (0.020) |
| **Vehicular Stops** | | | | |
| Donation | 0.043 | 0.019 | 0.005 | 0.017 |
|  | (0.031) | (0.017) | (0.007) | (0.016) |

* $p < 0.05$, ** $p < 0.01$, *** $p < 0.001$

Notes: We included 1,140 blocks, 620 blocks with a 7-Eleven and 512 control blocks. Aggregated in three-month time periods. We included 48 months of data, leaving all models with 18,240 observations. All models include block and time fixed effects. Clustered standard errors at the block level.

Figure 1 presents event study estimates comparing blocks with a 7-Eleven to major-intersection blocks across 16 three-month periods before and after the November 2017 donation. Consistent with the difference-in-difference estimates presented in Table 1, the results show a clear post-treatment increase in all investigatory stops -particularly pedestrian stops- at blocks with 7-Eleven. The top row reveals that, across all racial groups, the donation effect begins to rise immediately after November 2017 and becomes statistically significant by the second post-donation quarter. These increases persist and grow in magnitude over time, suggesting a sustained behavioral response by police rather than a temporary spike. The second row shows that the increase in stops is most pronounced for black individuals, who experienced a steep and steady rise in all investigatory stops and pedestrian stops after the donation. In contrast, the increases in whites and hispanics are smaller and less precise, with many post-donation coefficients overlapping zero and larger confidence intervals.

Figure 1: Event Study Analysis 7-Elevens vs Major Intersections

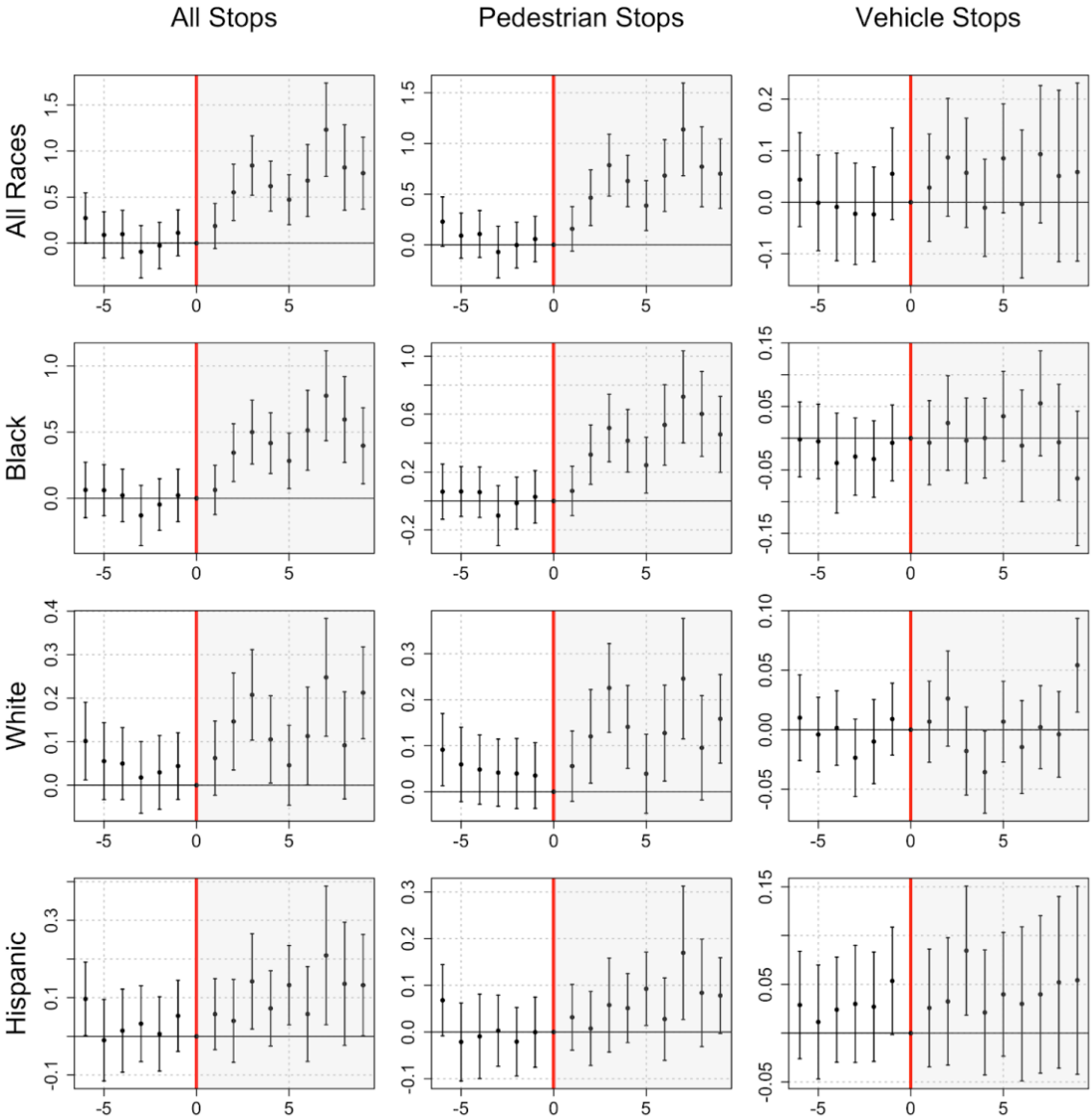

Notes: We included 1,140 blocks, 620 blocks with a 7-Eleven, and 512 control blocks. Aggregated in three-month time periods. We included 48 months of data, leaving all models with 18,240 observations. All models include block and time-fixed effects. Clustered standard errors at the block level. Confidence intervals at 95% level.

Across all panels, vehicle stops exhibit no meaningful change over time for any group, reinforcing that street-level pedestrian encounters drive the observed effects.

Table 2. Difference-in-Difference Results Other Convenience Stores vs Major Intersections

|  | All Races | Black | White | Hispanic |
|---|---|---|---|---|
| **All Investigatory Stops** | | | | |
| Placebo Treatment | 0.064 | 0.023 | 0.005 | 0.033 |
|  | (0.102) | (0.082) | (0.025) | (0.027) |
| **Pedestrian Stops** | | | | |
| Placebo Treatment | -0.056 | -0.056 | 0.007 | -0.008 |
|  | (0.092) | (0.073) | (0.024) | (0.023) |
| **Vehicular Stops** | | | | |
| Placebo Treatment | 0.120*** | 0.079** | -0.002 | 0.041*** |
|  | (0.029) | (0.025) | (0.004) | (0.011) |

\* p < 0.05, \*\* p < 0.01, \*\*\* p < 0.001

Notes: We included 4,259 blocks, 1,624 blocks with a convenience store, and 2,635 control blocks. Aggregated in three-month time periods. We included 48 months of data, leaving all models with 67,818 observations. All models include block and time-fixed effects. Clustered standard errors at the block level.

## Discussion

Our findings provide the first causal evidence that even modest private donations to police departments can alter the discretionary behavior of officers in ways that raise concerns about impartiality, transparency, and racial equity. By showing that a single $10,000 sponsorship to a police foundation gala led to hundreds of additional investigatory stops—particularly of Black pedestrians—in the vicinity of the donor's retail stores, this study highlights how seemingly benign acts of philanthropy can have disproportionate and racialized downstream consequences in public safety enforcement.

These findings advance our theoretical understanding of donor influence in policing by empirically validating the role of reciprocity bias. While previous literature has focused primarily on large-scale corruption or overt quid pro quo exchanges, our study demonstrates how behavioral shifts can emerge from legal, public forms of giving. This underscores a key insight from the psychology and economics of reciprocity: the size of the gift is not necessarily proportional to its influence. If officers respond measurably to a $10,000 donation, it is reasonable to hypothesize that the effects of far larger donations—many of which reach into the hundreds of thousands or even millions of dollars—may be even more significant and potentially more difficult to detect using existing public data.

Nonetheless, several limitations of our study merit attention. First, while our quasi-experimental design offers strong internal validity for the case of 7-Eleven's one-time donation, generalizing the effect to all donors or cities requires caution. Different corporate donors may enjoy varying degrees of visibility or salience among officers, and the spatial layout of a donor's locations may interact with patrol routes in complex ways. Second, while we detect a sustained behavioral response following the donation, we cannot observe officers' internal motivations or the exact mechanisms through which donor bias manifests—whether through departmental norms, direct instructions, or implicit psychological reciprocity. Although interviews of officers attending the TBG suggest they were "filled with gratitude" or "rejuvenated" because someone is "still backing the blue," we are unable to generalize this sentiment to every officer on the force ([FoP 2019](#)). These remain important areas for future research, especially for researchers with access to more granular data on officer assignments and attitudes.

Finally, this study raises broader normative questions about the governance of law enforcement. If donations—even small ones—can shift enforcement patterns, there is a pressing need to reconsider the role of private philanthropy in policing. As local governments face pressure to fund public safety without increasing taxes, police departments may become increasingly reliant on private donors. Our findings suggest that such reliance comes at a cost: the erosion of equitable, impartial policing. Any effort to regulate or reform police funding structures should therefore account not only for corruption risks, but also for the subtle, pervasive influence of reciprocity in shaping officer behavior.

**Materials and Methods**

To examine the effects of private donations on policing behavior, we leveraged the sponsorship of 7-Eleven to the Chicago Police Foundation's (CPF) True-Blue Gala in 2017. Using block-level data from Chicago between 2016 and 2020, we evaluate the effect of 7-Eleven donations on Chicago Police activity. To this end, we gathered data from five main sources.

Donations

Collecting donation data is complex since most police departments do not publicly disclose their donors and donation amounts. To overcome this challenge, we focus on CPF's public information about True Blue Gala (TBG) sponsorship. We collected TBG sponsorship data from 2015 to 2019 using archived versions of the CPF website through the Internet Archive's Wayback Machine and photographic materials posted by the CPF. TBG was canceled in 2020 due to COVID-19. Therefore, no sponsorship happened that year.

To isolate the effect of donations from other confounding factors, we focused on retail companies sponsoring TBG to leverage their physical presence across the city. Table 2 shows all the companies that have a physical presence in Chicago and that have sponsored the TGB between 2016 and 2024. We selected 7-Eleven 2017 donation for two reasons. First, it only sponsored the TBG one time during our period of study, allowing us to estimate the effect of one unique donation. Second, 7-Eleven has many stores across the city (N=116), which allows us to have enough statistical power to capture an effect.

Police Stop Data

Measuring police activity is often challenging since police departments do not make this information publicly available. To address this, we use police stop data as a proxy of police activity. We use data on investigatory stops released by the Chicago Police Department between 2016 and 2023. Investigatory stops involve temporary detentions and questioning of individuals based on "reasonable articulable suspicion" of criminal activity.

While direct data on patrol routes or deployment of officers are not publicly available, the number of stops in a given location offers a granular indicator of policing intensity when we account for criminal activity differences.

Crime Data

We leverage the geographic distribution of 7-Eleven stores across Chicago to construct a quasi-experimental setting. Since store locations are unlikely to be randomly distributed, we use block crime data published at the Chicago Data Portal to find a comparable group of blocks to isolate the effect of 7-Eleven donation on police investigatory stops.

Socio-Economic Controls

We leverage the geographic distribution of 7-Eleven stores across Chicago to construct a quasi-experimental setting. Since store locations are unlikely to be randomly distributed, we use data from the American Community Survey (ACS) five-year estimates (2016–2020) to identify a set of demographically and socioeconomically comparable blocks. Table 3 shows the differences between 7-Eleven blocks (N=628) and those with major intersections (N=512), suggesting that our two groups of blocks overall have no statistically significant mean differences

Table 3. Balance Table based on ACS 2016-2020 Characteristics

| Socio-Economic Controls | Control (N=512) Mean | Std. Dev. | Treatment (N=628) Mean | Std. Dev. | Diff. in Means | P-Value |
|---|---|---|---|---|---|---|
| Median Income | 100794.4 | 39014.9 | 101347.4 | 37996.9 | 553.0 | 0.811 |
| Share Black | 0.070 | 0.107 | 0.065 | 0.073 | -0.005 | 0.373 |
| Share White | 0.599 | 0.210 | 0.573 | 0.193 | -0.026 | 0.032 |
| Share Hispanic | 0.200 | 0.207 | 0.214 | 0.219 | 0.014 | 0.273 |
| Share Males between Ages 15-25 | 0.065 | 0.058 | 0.064 | 0.052 | -0.001 | 0.677 |
| Share College Graduates | 0.608 | 0.259 | 0.629 | 0.268 | 0.021 | 0.177 |
| Share Below Poverty Line | 0.104 | 0.078 | 0.103 | 0.067 | -0.000 | 0.918 |

<u>7-Eleven Store Location</u>

We obtained the location of 7-Eleven in Chicago through Data Axle, a commercial directory providing business listings by year. We filtered by city and used different names to identify a 7-Eleven store (e.g., "7-Eleven", "Seven Eleven", "Seven-Eleven"). We obtained a comprehensive list of store locations from 2010 onwards. In cases where only approximate location was available (e.g., ZIP code centroids), we successfully re-coded all entries using full address information to obtain precise coordinates.[6]

Figure 2 shows the distribution of 7-Eleven stores and major intersections across Chicago during our study period. Besides the geographic variation that 7-Eleven offers, we selected this one-time donation to separate potential effects related to long-term relationships between private donors and police departments from the donation effect.

---

[6] When a store appeared in two consecutive years but is missing in between, we assumed the location remained open.

Figure 2. Distribution of 7-Eleven and Major Intersections in Chicago, 2016-2020

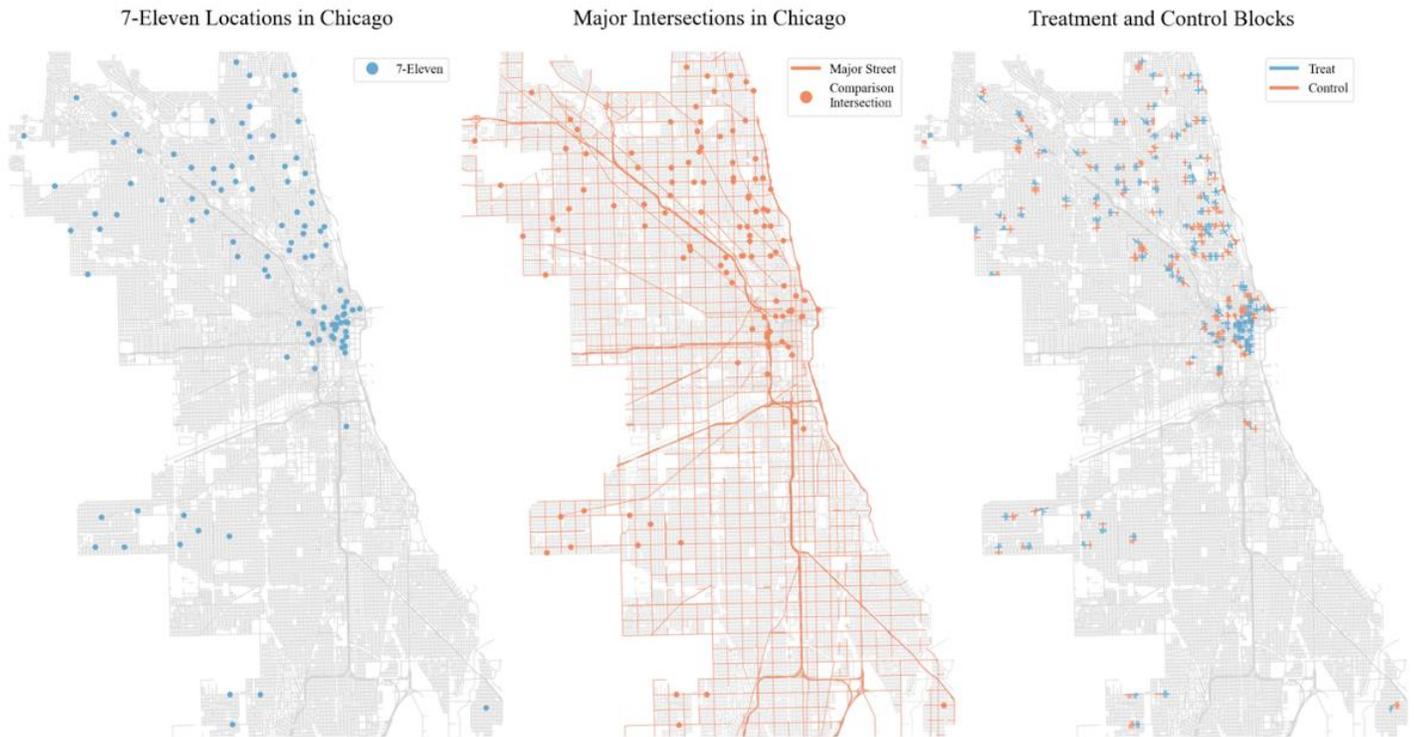

Empirical Strategy

Our analysis aims to estimate the effect of private donations to police departments on policing behavior. We measure private donations as 7-Eleven sponsorship of the 2017 Chicago Police TBG and police behavior as police investigatory stops in Chicago between 2016 and 2019. We preregistered our study after data collection and before estimating our models.[7] We employed a two-way fixed effects model comparing blocks within 330 feet of a 7-Eleven store (treatment group) to blocks within 330 feet of a major intersection (control group). Concretely, we estimated the following specification as our main model:

---

[7] Preregistration information is available at OSF: https://osf.io/ye9uk/. Initially, we proposed to compare 7-Eleven blocks against all other blocks without 7-Eleven, yet when we explored pre-trend differences, the best comparison group was blocks with major intersections. We also proposed in our preregistration study using 330 feet as block measure, yet when we estimated the monthly effect, our models did not have statistical power due to zero concentrations. The aggregate estimations are consistent with our estimations using 330 feet and are available upon request.

$$Police\ Stops_{it}^{S} = \beta_0 + \beta_1 Donation_{it} + \alpha_i + \delta_t + \varepsilon_{it} \quad (1)$$

where *s* represents the three measures of police investigatory stops by race in a block i in period t, where t has a periodicity of three months. We estimate equation (1) using three outcome variables: (i) all investigatory police stops, (ii) pedestrian investigatory police stops, and (iii) vehicular investigatory police stops. In our preregistration, we hypothesized that the donation effect should only be observable in pedestrian stops, which would drive an effect in all investigatory stops. While a donation could increase police activity around the retail store in general, we hypothesized that police officers would only change their behavior to increase retailer protection, and a vehicular stop is unlikely to be related to retail crime.

Our donation variable is represented by a dummy variable that takes the value of one when a block has a 7-Eleven and the observation is after November 2017, when the sponsorship to the TBG happened, and zero otherwise.[8] We included block ($\alpha_i$) and time ($\delta_t$) fixed effects to account for time-invariant block characteristics and seasonal temporal changes. Finally, we included cluster standard errors at the block level ($\varepsilon_{it}$). We included in our appendix other versions of our main specification controlling for crime, showing our results are robust to changes in our model.

We estimate an event study specification with block and time fixed effects to examine the dynamic relationship between 7-Eleven donation and police stops. This approach serves two purposes. First, it allows us to assess the plausibility of the parallel trends assumption before November 2017. Second, the model captures how police stops evolved after the 7-Eleven donation, enabling us to estimate period-specific effects and explore temporal dynamics of the donation's impact. Specifically, we estimated the following model:

---

[8] We aggregated t every three months using November 2017 as our indicator to aggregate.

$$Police\ Stops_{it}^s = \beta_0 + \beta_1 \sum_{k=-m}^{-1} Donation_{it}^k + \beta_2 \sum_{k=1}^{n} Donation_{it}^k + \alpha_i + \delta_t + \varepsilon_{it} \quad (2)$$

where $Donation_{it}^k$ is an event-time indicator equal to one if the observation is k periods away from the donation event (with k=0 omitted as the reference period); $\beta_1$ captures the average difference in the outcome during the pre-donation period (from k= -m to k= -1, with m=2,…,6), and $\beta_2$ captures the average difference in the post-donation period (from k=1 to k=n, where n=2,…,9). We estimated an additional specification of model (2) to assess crime pre-trends (see figure 1A in the Appendix).

We estimated three tests replicating our main specification to assess the likelihood of obtaining false statistically significant results (1). First, we changed our treated group from blocks with a 7-Eleven to blocks with other convenience stores and maintained our control group (Table 3). Second, we replicated our analysis during the pre-treatment period, recoding our donation dummy variable to take the value of one from November 2016, and excluding all periods after November 2017 (see table 3A in the Appendix). Finally, we simulated different versions of our treatment, changing and maintaining November 2017 as the donation date. We randomly assigned Chicago blocks to a treatment and control group and estimated our model (1). We repeated this random assignment and estimation 100 times (see figure 3A in the Appendix). We could not replicate our results with these three tests, suggesting that our estimations are not false significant effects.

**Appendix**

Donations for TBG (2016-2024)

Table 1A shows our donation data information showing all private corporations with physical presence across Chicago and whether they have sponsored the TBG between 2016 and 2024, excluding 2020 due to the Covid-19 pandemic.

Table 1A. Chicago Police Foundation Gala Sponsors by Year

| Company | 2016 | 2017 | 2018 | 2019 | 2021 | 2023 | 2024 |
|---|---|---|---|---|---|---|---|
| @ Properties | No | Yes | Yes | Yes | Yes | No | No |
| 7Eleven | No | Yes | No | No | No | No | No |
| 8 Hospitality | No | No | No | No | Yes | Yes | Yes |
| AED Professionals | No | No | No | No | Yes | Yes | Yes |
| Allan Reich | No | No | No | No | No | Yes | Yes |
| American Airlines | No | No | No | Yes | No | No | No |
| American Heritage Protective Services | No | No | No | No | No | No | Yes |
| Axon | No | Yes | Yes | Yes | Yes | No | No |
| BMO Harris Bank | Yes | Yes | Yes | Yes | No | No | No |
| Cigna | Yes | Yes | Yes | Yes | No | No | No |
| CNA | No | No | No | Yes | Yes | No | No |
| Columbia Pipe and Supply | No | No | Yes | Yes | No | No | No |
| ComEd | Yes | Yes | Yes | No | No | No | No |
| Daniel Micic | No | No | No | No | No | No | Yes |
| Dr. Scholls | No | No | No | No | No | No | Yes |
| Duchossois Capital Management | No | No | No | No | No | No | Yes |
| Eli's Cheesecake | No | Yes | No | No | No | Yes | Yes |
| Employco USA | No | No | No | No | No | No | Yes |
| Energy Distribution Partners | No | No | No | No | No | No | Yes |
| Fifth Third Bank | No | No | No | No | No | Yes | Yes |
| Forward Together | No | No | No | No | Yes | No | No |
| Four Seasons | No | No | No | No | No | Yes | Yes |
| Gaslight Studios | No | No | No | No | No | No | Yes |
| GCM Grosvenor | No | Yes | No | No | No | No | No |
| GPG Strategies | No | No | Yes | Yes | No | No | No |
| Greeley and Hansen | Yes | Yes | Yes | Yes | Yes | No | No |
| Hilco Global | Yes | No | No | No | No | No | No |
| Horizon (Therapeutics) | No | No | No | No | No | Yes | No |
| Hyatt Regency | No | Yes | No | No | No | No | No |
| Jim Kallas Calumet Hospitality | No | No | No | No | Yes | No | No |
| John Robak | No | No | No | No | No | Yes | Yes |
| JP Morgan Chase | No | No | No | No | No | No | Yes |
| Justine Fedak Corporate Hippie | No | No | No | No | Yes | No | No |
| Kevin and Sarah Bruno | No | No | No | No | No | No | Yes |
| Landis Family Foundation | No | No | No | No | No | Yes | Yes |
| Law Offices of Debra Dimaggio | No | No | No | No | Yes | Yes | No |
| LAZ Parking | Yes | Yes | Yes | Yes | Yes | Yes | No |
| Louis D'Angelo | No | No | No | No | No | No | Yes |
| Magellan Coporation | No | Yes | Yes | Yes | No | No | No |
| Maverick Hotel and Restaurants | No | No | No | No | No | Yes | Yes |
| MB Financial | No | Yes | Yes | No | No | No | No |
| Medical Shipment | No | No | No | No | Yes | No | No |
| Modern Luxury CS | Yes | Yes | No | No | No | No | No |
| Motorolla Solutions | No | No | No | Yes | No | No | No |
| Noahs Ark | No | No | No | No | Yes | No | No |
| OverNorth | No | No | Yes | No | No | No | No |
| Pamela and Alfredo Capitanini | No | No | No | No | No | Yes | Yes |
| Peoples Gas | No | No | No | No | No | Yes | Yes |
| PNCBank | No | No | Yes | No | Yes | No | No |
| Rainbow Sandals | No | No | No | No | Yes | No | No |
| Richard Gamble | No | No | No | No | No | No | Yes |
| Risk Management Solutions of America | No | No | No | No | No | No | Yes |
| RME (Rubino and Mesia Engineers) | No | No | No | No | Yes | No | No |
| RSM (RSM International) | No | No | No | No | Yes | No | No |
| Shotspotter | No | No | Yes | No | Yes | No | Yes |
| Sudler | No | No | Yes | No | No | No | No |
| Supreme Lobster | No | No | No | No | No | No | Yes |
| Taft Law | No | No | No | No | No | No | Yes |
| Tamar Productions | No | Yes | No | No | No | Yes | Yes |
| The Horton Group | No | No | No | No | No | Yes | No |
| The Sotos Law Firm | No | No | No | No | Yes | No | No |
| Third Coast Hospitality | No | No | No | No | Yes | No | No |
| Time Zone One | No | No | No | No | Yes | No | No |
| Torshen Capital Management/Kay Torshen Foundation | No | No | Yes | Yes | No | No | No |
| Tritech/Central Square | No | No | No | Yes | Yes | No | No |
| Tullman Community Ventures | No | No | No | No | No | Yes | No |
| United Airlines | No | No | No | Yes | No | No | No |
| United Scrap | No | No | No | No | No | Yes | Yes |
| United Service Companies | Yes | Yes | Yes | Yes | Yes | Yes | Yes |
| US Bank | No | Yes | Yes | Yes | No | No | No |
| USI | Yes | Yes | Yes | Yes | Yes | No | Yes |
| Verizon | No | No | No | Yes | Yes | Yes | Yes |
| Weber Shandwick | Yes | Yes | Yes | Yes | No | No | No |
| William Blair | No | No | No | No | No | No | Yes |
| Wintrust | No | No | No | No | No | No | Yes |
| Wilson Companies | No | No | No | No | No | Yes | No |

Control Group Blocks

We compared 7-Eleven blocks to three groups of blocks to identify a group of blocks with similar crime dynamics to rule out potential effects in police stops due to crime dynamics. We compared (i) 7-Eleven blocks to all other blocks without a 7-Eleven, (ii) 7-Eleven blocks to all other blocks with a convenience store, and (iii) 7-Eleven blocks to all other blocks with major intersections. Figure 1A shows that the best comparison group is blocks with major intersections since these have no statistically significant difference in crime rates before the donation occurred.

Figure 1A. Crime Pre-trends Potential Control Blocks

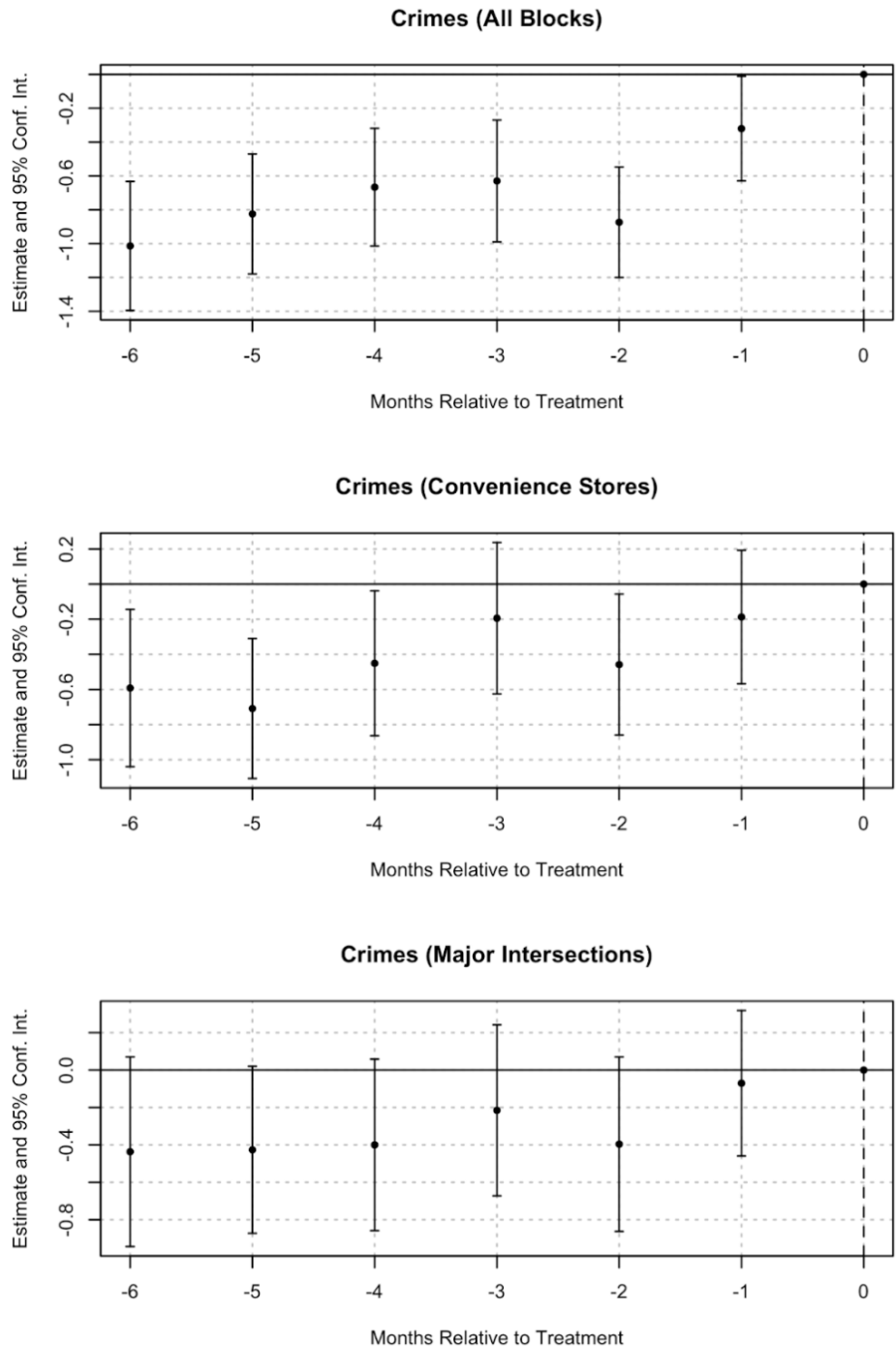

Notes: For all models there were 628 treated blocks and 21 months of data aggregated in 3-month periods. For the all block comparison, there were 33,162 comparison blocks for a total of 236,530 observations. For other convenience stores there were 2,669 comparison blocks for a total of 20,947 observations. For the major intersection comparison there were 512 comparison blocks for a total of 7,980 observations.

Once we found a group of blocks with comparable crime trends, we checked potential differences across other block socio-economic characteristics. In our main table, Table 3, we present that based on the ACS from 2016 to 2020, there were no statistically significant differences between 7-Eleven and major intersection blocks. Table 2A shows a similar trend for the years 2011 to 2015, confirming that these blocks had balanced characteristics during our study period, but these similarities have been constant over time.

Table 2A. Balance Table based on ACS 2011-2015 Characteristics

|  | Control (N=495) | | Treatment (N=628) | | Diff. in Means | P-value |
|---|---|---|---|---|---|---|
|  | Mean | Std. Dev. | Mean | Std. Dev. | | |
| Median Income | 80964.3 | 30103.0 | 78277.1 | 29434.0 | -2687.2 | 0.136 |
| Share Black | 0.076 | 0.094 | 0.070 | 0.080 | -0.005 | 0.328 |
| Share White | 0.619 | 0.201 | 0.588 | 0.195 | -0.031** | 0.009 |
| Share Hispanic | 0.186 | 0.200 | 0.208 | 0.218 | 0.022+ | 0.075 |
| Share Males between Ages 15 and 25 | 0.071 | 0.065 | 0.069 | 0.053 | -0.002 | 0.607 |
| Share College Graduates | 0.616 | 0.248 | 0.584 | 0.265 | -0.031* | 0.043 |
| Share Below Poverty Line | 0.135 | 0.096 | 0.135 | 0.081 | 0.000 | 0.983 |

Robustness Checks

We estimated alternative versions of our main specification (Table 1) to assess the robustness of our findings. In table 3A, we included lagged block-level crime as a control to account for the possibility that changes in police stops may be driven by contemporaneous or recent criminal activity rather than the donation itself. The inclusion of this variable does not substantially alter the estimated treatment effects, suggesting that our results are not confounded by shifts in local crime rates.

Table 3A: Difference-in-Difference Results 7-Elevens vs Major Intersections with Crime

|  | All Races | Black | White | Hispanic |
|---|---|---|---|---|
| **All Investigatory Stops** | | | | |
| Donation | 0.597*** | 0.416*** | 0.091*** | 0.078** |
|  | (0.127) | (0.096) | (0.025) | (0.028) |
| **Pedestrian Stops** | | | | |
| Donation | 0.555*** | 0.399*** | 0.086*** | 0.062** |
|  | (0.114) | (0.090) | (0.024) | (0.020) |
| **Vehicular Stops** | | | | |
| Donation | 0.042 | 0.018 | 0.005 | 0.017 |
|  | (0.030) | (0.016) | (0.007) | (0.016) |

* $p < 0.05$, ** $p < 0.01$, *** $p < 0.001$

Notes: We included 1,140 blocks, 628 blocks with a 7-Eleven and 512 control blocks. Aggregated in three-month time periods. We included 48 months of data, leaving all models with 18,240 observations. All models include lagged crime as control, block and time fixed effects. Clustered standard errors at the block level.

To assess the temporal dynamics of the treatment effect more comprehensively, we re-estimated our event study model presented in Figure 1 using a longer time window before and after the 7-Eleven donation. As shown in figure 2A, the results are consistent with our main specification. We observe a clear post-donation increase in all stops and pedestrian stops, particularly among blacks. This extended window helps to rule out the possibility that short-term fluctuations or pre-existing trends drive our findings.

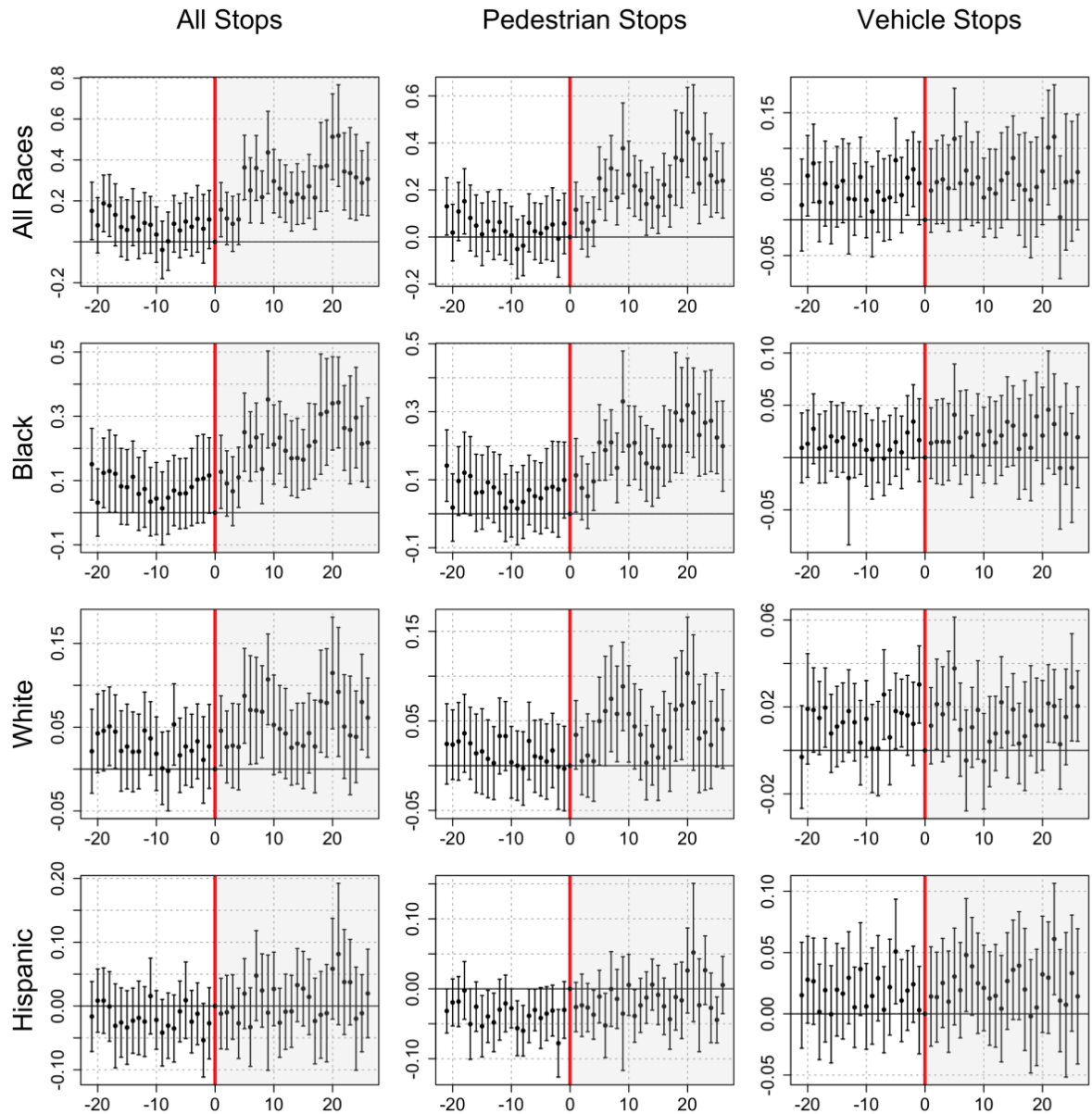

Figure 2A: Event Study Analysis 7-Elevens vs Major Intersections

Notes: We included 1,140 blocks, 628 blocks with a 7-Eleven, and 512 control blocks. We included 48 months of data, leaving all models with 54,720 observations. All models include block and time-fixed effects. Clustered standard errors at the block level. Confidence intervals at 95% level.

Additional Placebo Checks

As an additional placebo test, we assigned a false treatment date prior to the actual November 2017 7-Eleven donation and re-estimated our difference-in-difference specification on pre-treatment data only. Table 3A reports the estimated placebo treatment effect across all investigatory, pedestrian, and vehicular stops, disaggregated by racial group. Across all models, the placebo treatment effects have a different direction, magnitude, and are not statistically significant. This finding supports the plausibility of parallel trends prior to the actual intervention.

Table 3A: Fixed Effects with Placebo Treatment in Pre-Period

|  | All Races | Black | White | Hispanic |
| --- | --- | --- | --- | --- |
| **All Investigatory Stops** | | | | |
| Placebo | -0.012 | 0.017 | -0.029 | -0.002 |
|  | (0.092) | (0.076) | (0.027) | (0.029) |
| **Pedestrian Stops** | | | | |
| Placebo | -0.042 | -0.001 | -0.038 | -0.004 |
|  | (0.087) | (0.073) | (0.025) | (0.023) |
| **Vehicular Stops** | | | | |
| Placebo | 0.030 | 0.018 | 0.010 | 0.002 |
|  | (0.027) | (0.017) | (0.009) | (0.018) |

\* $p < 0.05$, \*\* $p < 0.01$, \*\*\* $p < 0.001$

Notes: We included 1,140 blocks, 628 blocks with a 7-Eleven and 512 control blocks. Placebo is defined in November 2016. Aggregated in three-month time periods. We included 21 months of data, leaving all models with 7,980 observations. All models include block and time fixed effects. Clustered standard errors at the block level.

As a final placebo test, we randomly reassigned the donation indicator (blocks with 7-Eleven) to 100 groups of blocks while holding all other aspects of the data and model constant. Figure 3A shows the distribution of estimated treatment effects for all stops and pedestrian stops across these 100 placebo replications separately by racial group. The vertical red line represents the actual treatment effect from our main model presented in Table 1. In each case, the observed donation effect lies at or beyond the upper tail of the distribution of placebo estimates, indicating that the observed effect is unlikely to have arisen by chance.

Figure 3A. Placebo Distribution of Estimated Treatment Effects from 100 Random Assignments

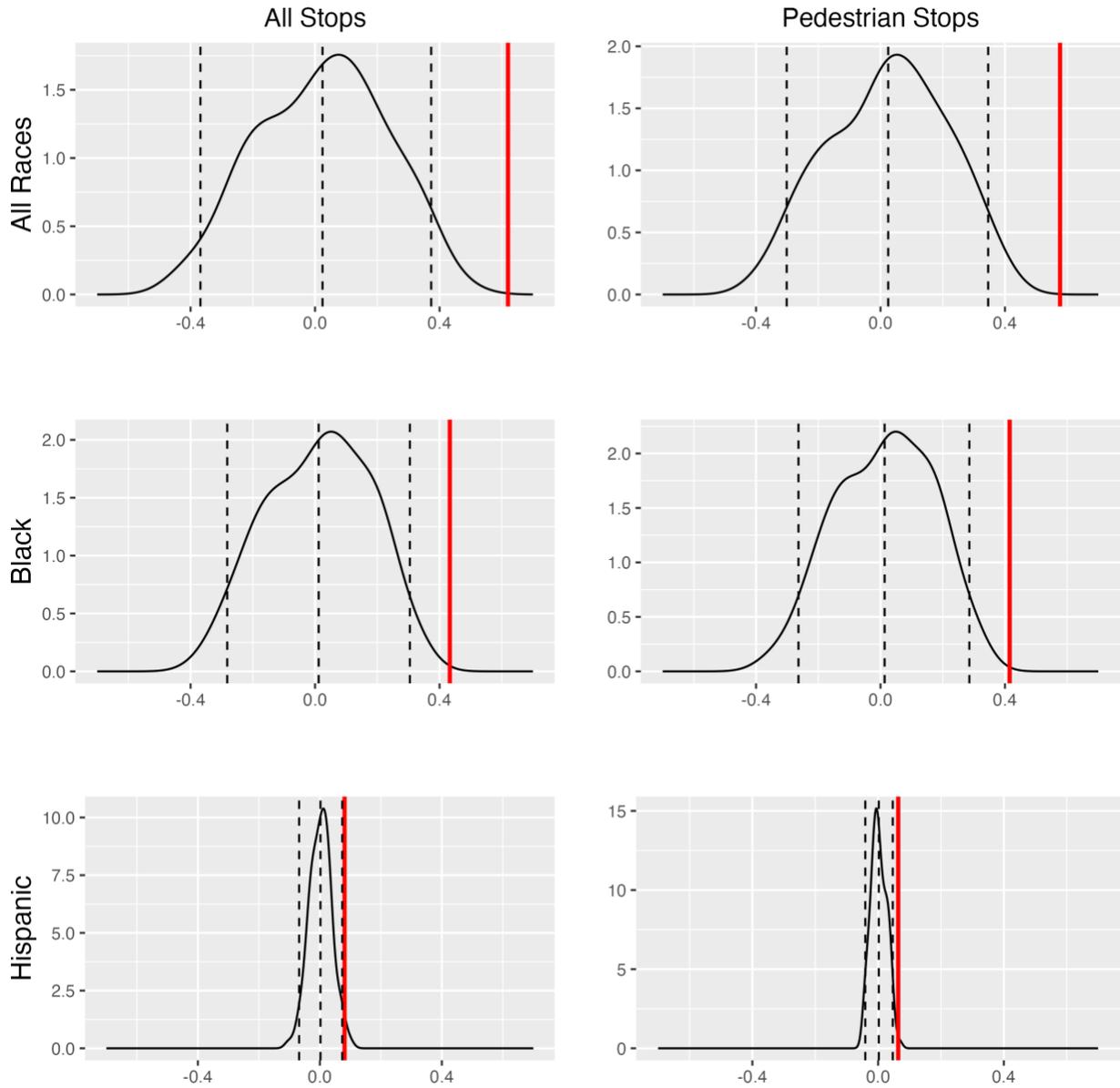

Note: We estimated 100 times our main specification (Equation 1) using a random treatment (donation). The black line shows the distribution of the coefficients obtained from the 100 estimations, and the red line represents our true coefficient.